\documentclass[final,1p,times]{elsarticle} 

\usepackage{graphicx}
\usepackage{amssymb} 
\usepackage{amsthm} 
\usepackage{lineno}

\journal{Nuclear Physics A} 

\begin{document}

\begin{frontmatter} 

\title{Initial state fluctuations and higher harmonic flow\\ in heavy-ion collisions }

\author[auth2]{Charles Gale}
\author[auth2]{Sangyong Jeon}
\author[auth1]{Bj\"orn Schenke}
\author[auth3]{Prithwish Tribedy}
\author[auth1]{Raju Venugopalan}
\address[auth2]{Department of Physics, McGill University, 3600 University Street, Montreal, Quebec, H3A\,2T8, Canada}
\address[auth1]{Physics Department, Brookhaven National Laboratory, Upton, NY 11973, USA}
\address[auth3]{Variable Energy Cyclotron Centre, 1/AF Bidhan Nagar, Kolkata 700064, India}

\begin{abstract} 
A framework combining Yang-Mills dynamics of the pre-equilibrium glasma with relativistic viscous hydrodynamic evolution of the quark-gluon plasma and hadron gas phases is presented. Event-by-event fluctuations of nucleon positions and color charges are taken into account, leading to negative binomial fluctuations of gluon multiplicities. Experimental anisotropic flow coefficients $v_2$-$v_5$ of charged hadron distributions in heavy-ion collisions at the Large Hadron Collider are well described. Furthermore, event-by-event distributions of $v_2$, $v_3$ and $v_4$ measured by the ATLAS collaboration are reproduced.
\end{abstract} 

\end{frontmatter} 


\section{Introduction}
Heavy-ion collisions at the Relativistic Heavy Ion Collider (RHIC) and the Large Hadron Collider (LHC) allow for systematic exploration of the high temperature many-body dynamics of a non-Abelian quantum field theory. 
The study of multiplicity fluctuations and anisotropic flow harmonics $v_n$ at both RHIC \cite{Adare:2011tg,Sorensen:2011fb} and LHC \cite{ALICE:2011ab,ATLAS:2012at,CMS:2011} potentially allows to obtain information on the strongly correlated non-equilibrium glasma regime and the transport properties of the nearly equilibrated quark-gluon plasma and hadron gas phases. To do so, a sophisticated theoretical description of all stages of the evolution of the complex system created in heavy-ion collisions is needed. We present a promising theoretical framework for this task with the combination of the IP-Glasma model with relativistic viscous hydrodynamics.

\section{IP-Glasma model}
The IP-Glasma model \cite{Schenke:2012wb,Schenke:2012hg} combines the IP-Sat (Impact Parameter Saturation) 
model~\cite{Bartels:2002cj,Kowalski:2003hm} of high energy nucleon and nuclear wavefunctions with the classical 
Yang-Mills (CYM) dynamics of the glasma fields produced after the heavy-ion collision~\cite{Kovner:1995ja,Kovchegov:1997ke,Krasnitz:1999wc,Krasnitz:2000gz,Lappi:2003bi}. It relates the deeply inelastic scattering (DIS) constrained nuclear dipole cross-sections to the initial classical dynamics of highly occupied gluon ``glasma'' fields after the nuclear collision. Given an initial distribution of color charges in the high energy nuclear wavefunctions, the IP-Glasma framework computes the strong early time multiple scattering of gluon fields by event-by-event solutions of Yang-Mills equations.

The IP-Glasma model naturally includes the effect of several sources of quantum fluctuations that can influence hydrodynamic flow: fluctuating distributions of nucleons in the nuclear wavefunctions and intrinsic fluctuations of the color charge distribution.
This results in ``lumpy'' transverse projections of the gluon field configurations that vary event to event. The scale of this lumpiness is given on average by the inverse nuclear saturation scale $Q_s$ which corresponds to distance scales smaller than the nucleon size.
The details of the implementation can be found in \cite{Schenke:2012wb,Schenke:2012hg}.

\section{Multiplicities}
\begin{figure}[tb]
  \begin{center}
    \begin{minipage}{0.495\textwidth}
    \includegraphics[width=7cm]{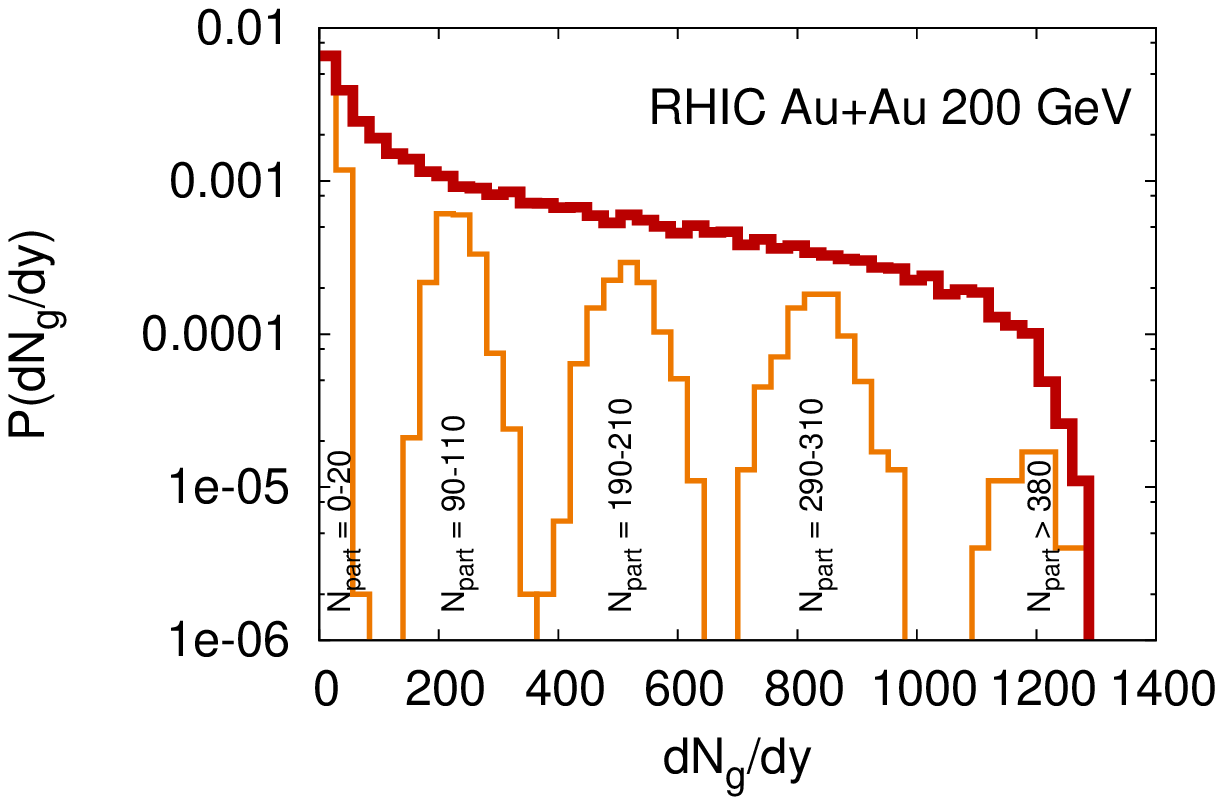} 
    \end{minipage}
    \hfill
    \begin{minipage}{0.495\textwidth}
    \includegraphics[width=7cm]{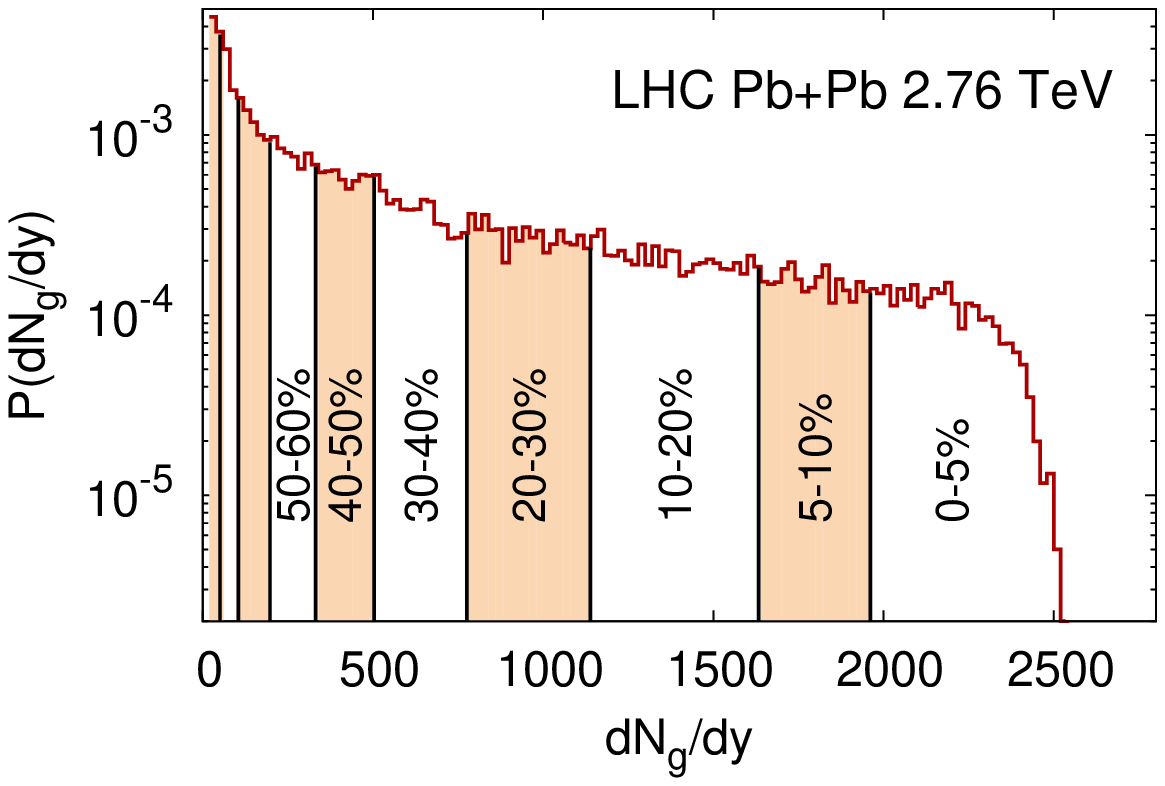}
    \end{minipage}
    \caption{Probability distributions of gluon multiplicities $dN_g/dy$.
      Left: RHIC. Shown are also the distributions for some limited ranges of $N_{\rm part}$, which are described by negative binomial distributions.
      Right: LHC with centrality classes.
      \label{fig:dNdyweighted}}
  \end{center}
\end{figure}

In Fig.\,\ref{fig:dNdyweighted} we present the probability distribution of $dN_g/dy$ at RHIC and LHC energies. An essential ingredient is the probability distribution of impact parameters, which is determined by the Glauber model. One could in principle compute this distribution in the Glasma framework, but a first principles computation is extremely difficult (see \cite{Schenke:2012hg}). 
For RHIC we also show five distributions obtained by constraining the number of participants to demonstrate that their shape resembles negative binomial distributions. For LHC we indicate ranges for different centrality classes.

\section{Transition to hydrodynamics}
When we switch from the CYM description to hydrodynamics we construct the fluid's initial energy momentum tensor $T^{\mu\nu}_{\rm fluid} = (\epsilon + {\cal P})u^\mu u^\nu - {\cal P}g^{\mu\nu} + \Pi^{\mu\nu}$ from the energy density in the fluid's rest frame $\varepsilon$, the flow velocity $u^\mu$, and, using an equation of state, the local pressure ${\cal P}$ at each transverse position. $\varepsilon$ and $u^\mu$ are obtained by solving $u_\mu T^{\mu\nu}_{\rm CYM} = \varepsilon u^\nu$, using the fact that $u^\mu$ is a time-like eigenvector of $T^{\mu\nu}_{\rm CYM}$ and satisfies $u^2 =1$. 
For the present study we do not extract the very non-equilibrium $\Pi^{\mu\nu}$ but assume an efficient instability driven isotropization of the system and
set the initial $\Pi^{\mu\nu}$ to zero. A 3+1 dimensional Yang-Mills simulation including quantum fluctuations can hopefully provide the full dynamics for this mechanism in the future. Further details on the hydrodynamic simulation can be found in \cite{Gale:2012rq}.

We determine centrality classes using the multiplicity distribution of gluons (Fig.\,\ref{fig:dNdyweighted} right)
much alike the procedure followed by the heavy-ion experiments using charged particle multiplicity distributions.

The hydrodynamic stage, including a Cooper-Frye freeze-out and resonance decays, is simulated using \textsc{music} \cite{Schenke:2010nt,Schenke:2010rr,Schenke:2011tv,Schenke:2011bn}.
Using the final particle distributions we determine 
\begin{equation}
  v_n(p_T) =\langle \cos(n(\phi-\psi_n)) \rangle 
\end{equation}
in every event using the exact event plane
\begin{equation}
  \psi_n=\frac{1}{n}\arctan\frac{\langle \sin(n\phi)\rangle}{\langle \cos(n\phi)\rangle}\,.
\end{equation}
where $\langle \cdot \rangle$ is the average over the smooth particle distribution function.

\section{Flow results}
Calculated root-mean-square (rms) $v_n(p_T)$ for $10-20\%$ central and more peripheral collisions agree very well \cite{Gale:2012rq} with experimental data from the ATLAS collaboration \cite{ATLAS:2012at}, which are determined using the experimental event-plane method that agrees well with the rms values \cite{ATLAS:2012Jia,Jia:2012ve}.
Integrated rms $v_2$, $v_3$ and $v_4$ also agree very well \cite{Gale:2012rq} with available $v_n\{2\}$ results (obtained using two-particle correlations) from the ALICE collaboration \cite{ALICE:2011ab}.

\begin{figure}[htb]
  \vspace{-0.5cm}
   \begin{center}
    \begin{minipage}{0.495\textwidth}
     \includegraphics[width=7cm]{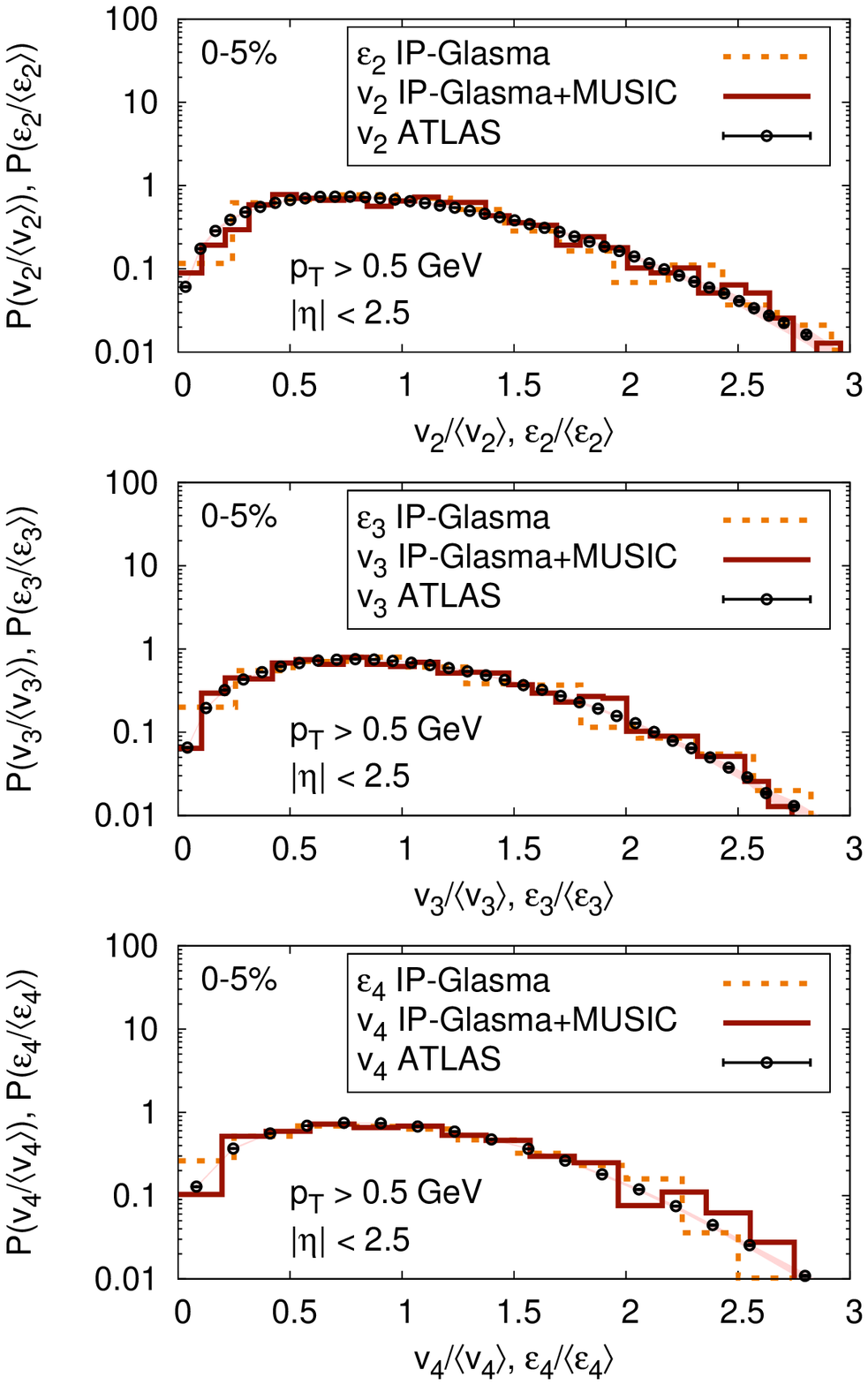}
    \end{minipage}
    \hfill
    \begin{minipage}{0.495\textwidth}
    \includegraphics[width=7cm]{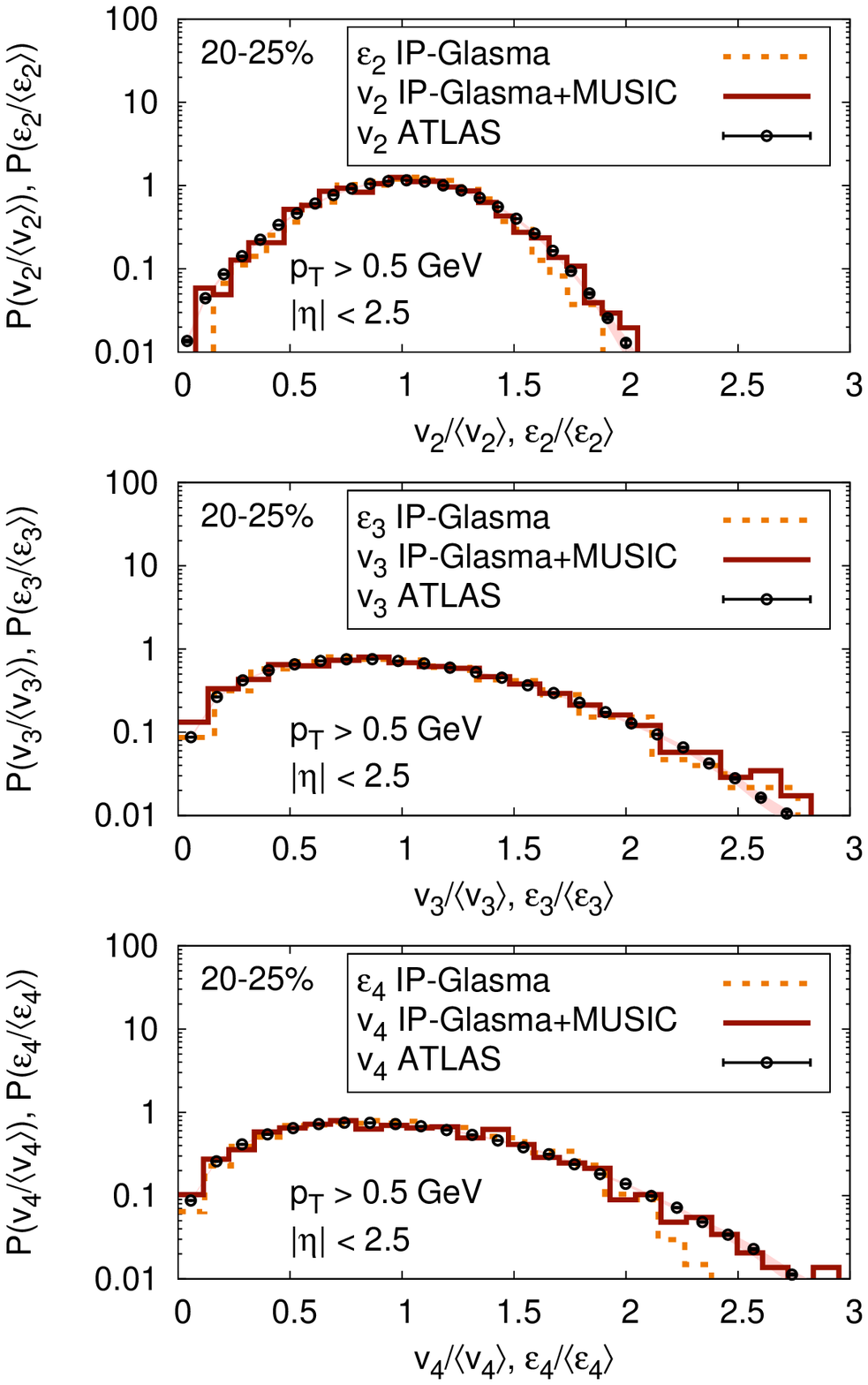}
    \end{minipage}
     \vspace{-0.5cm}
     \caption{Scaled distributions of $v_2$, $v_3$ and $v_4$ as well as $\varepsilon_2$, $\varepsilon_3$ and $\varepsilon_4$ compared to experimental data from the ATLAS
     collaboration \cite{ATLAS:2012Jia,Jia:2012ve}. Using 750 (0-5\%) and 1300 (20-25\%) events. Bands are systematic experimental errors.}
     \label{fig:vnenDist-20-25}
   \end{center}
   \vspace{-0.5cm}
\end{figure}

Here, we present event-by-event distributions of $v_2$, $v_3$ and $v_4$ for two centrality classes
 compared to results from the ATLAS collaboration \cite{ATLAS:2012Jia,Jia:2012ve} in Fig.\,\ref{fig:vnenDist-20-25}.
Distributions are scaled by their mean to be able to compare to initial eccentricity distributions at the same time.
We find that the initial eccentricity distributions are a good approximation to the distribution of experimental $v_n$.
Exact agreement with the data in the 20-25\% central bin, however, is only achieved with the $v_n$ distributions obtained after hydrodynamic evolution.

\section{Summary}

The IP-Glasma+\textsc{music} model produces anisotropic flow in excellent agreement with experimental data from the LHC. 
It is particularly remarkable that the event-by-event distributions of $v_n$ agree with experimental results from the ATLAS collaboration,
indicating that the fluctuations in the initial state are described accurately by the model.
The successful description of a wide range of data within this model provides a framework to precisely determine key aspects of the complex dynamics of heavy ion collisions.

\section*{Acknowledgments}
BPS\ and RV\ are supported under DOE Contract No.DE-AC02-98CH10886 and acknowledge support from a BNL Lab Directed Research and Development grant.
CG and SJ are supported by the Natural Sciences and Engineering Research Council of Canada.
We gratefully acknowledge computer time on the Guillimin cluster at the CLUMEQ HPC centre, a part of Compute Canada HPC facilities.
BPS gratefully acknowledges a Goldhaber Distinguished Fellowship from Brookhaven Science Associates.

\bibliographystyle{h-elsevier}
\bibliography{spires}

\end{document}